\documentclass[journal]{IEEEtran}

\ifCLASSINFOpdf
\else
\fi

\usepackage{verbatim}
\usepackage{graphicx} 
\usepackage{subfigure}
\usepackage{caption}
\usepackage{url}
\begin{document}

\title{Towards Zero Touch Networks: From the Perspective of Hierarchical Language Systems}

\author{Guozhi Lin, Jingguo Ge, and Yulei Wu, \textit{Senior Member, IEEE}
% <-this % stops a space
\thanks{This work was supported in part by the Special project for innovative methods of the Ministry of science and technology of China under Grant  2019IM020100.}
\thanks{G. Lin and J. Ge are with Institute of Information Engineering, Chinese Academy of Sciences, Beijing, 100093, China and School of Cyber Security, University of Chinese Academy of Sciences, Beijing, 100049, China.}% <-this % stops a space
\thanks{Y. Wu is with College of Engineering, Mathematics and Physical Sciences, University of Exeter, Exeter, EX4 4QF, UK.}% <-this % stops a space
\thanks{Corresponding author: Yulei Wu (Y.L.Wu@exeter.ac.uk)}

}

\maketitle

\begin{abstract}
With ever-increasing complexity and dynamicity of communication networks, intelligent network operation and maintenance has become more important to network operators. With the fast development of artificial intelligence, concepts such as ``Zero Touch'', ``Intent-based'', ``Knowledge-defined'' and ``Self-driving'' networks have become well-known in networking community for a great vision of making networks automatically manageable and responsive to user demands. 
This article discusses how to achieve Zero Touch Networks from the perspective of language-like systems. We propose a novel hierarchical `language' framework dedicated for networks to enable the Zero Touch Network, which covers from symbolizing network components, a unified framework for understanding network systems, to the logical description with network semantics. A case study based on the proposed language framework is provided.
Finally, we discuss the challenges and open issues of intelligence models  for zero touch networks. 
\end{abstract}


\IEEEpeerreviewmaketitle

\section{Introduction}
\IEEEPARstart
In recent years, the proliferation of diversified services, e.g., augmented reality/virtual reality (AR/VR), and emerging network technologies, e.g., 5G/B5G and network digital twins~\cite{9795043}, have resulted in a complex and dynamic networking system. Manually managing communication networks is an unacceptable option for network operators. Network autonomy is a clear trend in future that has attracted tremendous attention in the networking community. The ``Zero Touch Network'' \cite{9749186} which is to fully integrate the processing of perception, control, adaptivity and assurance to realize the automated management of networks without human intervention, is facing diversified discussions.

Thanks to the great progress made by machine learning, especially deep learning in automating tasks such as protein folding problems~\cite{0Improved} and autonomous driving~\cite{9583858}, many machine learning methods have also been explored and applied to network automation~\cite{9200928}, including resource scheduling~\cite{9846947}, anomaly detection~\cite{9851464}, proactive defense~\cite{9795042}, etc. This provides an opportunity for networks to perform intelligently when facing complex tasks. Despite of the powerful capability
of machine learning models on network automation in certain
scenarios, they are still designed based on a priori human knowledge and representation of specific tasks with little consideration on the form of intelligence required  for zero touch management. This is mainly due to the complexity of network systems which are difficult to generalize the essential properties of network intelligence. Attempting the solution of this challenging problem is an important boost in the development of zero touch networks.

\begin{figure}[htb]
	\centering
	\includegraphics[width=1\linewidth]{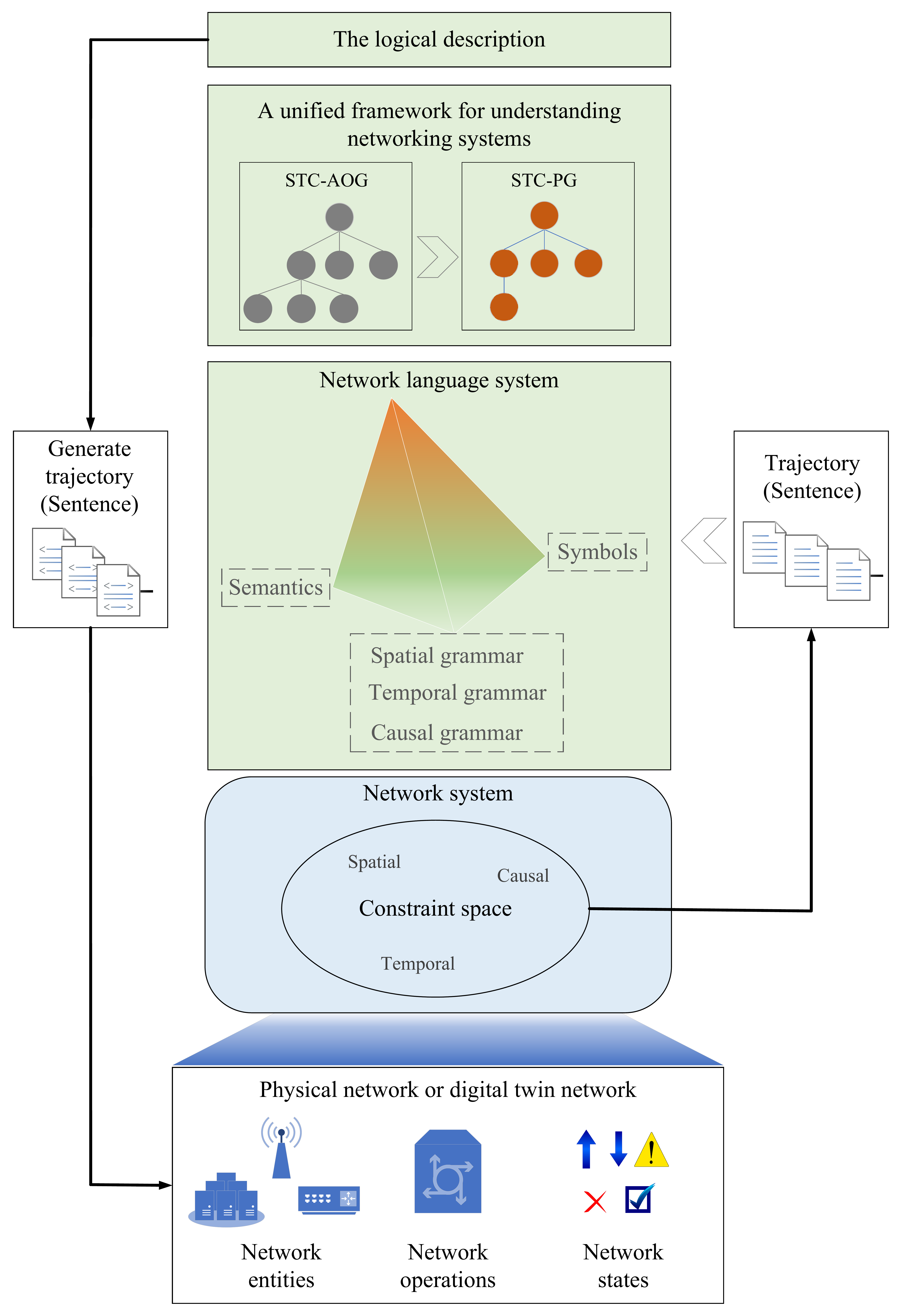}
	\caption{The zero touch network from the perspective of language-like systems. }
	\label{fig:Program overview}
\end{figure}

Fundamentally, the premise of realizing a dynamic system that can operate with zero touch is that, there is a mechanism to ensure its self-running behaviors within the bounds of constraints.
The behavior trajectories (or phenomena) of the system can be represented explicitly in the form of a language, such as a symbol string marking behavior. Language is a form of external manifestation of the constraints shown inherently in the system. Analogous to the human natural language, the position and semantics of each symbol is constrained by contextual information inherently embedded in human communication. Compared with natural language systems, the form of constraints is more complex and diverse in networking systems.  If there is a language system that can be induced by the internal constraints of networking systems, the trajectories (or phenomena) expressed by the language can be seamlessly integrated into the networking system  to achieve self-control and management, as shown in the flow depicted by the solid black lines with arrows in Fig. \ref{fig:Program overview}.

With the ultimate goal of realizing zero touch networks to
autonomously manage the network, three key constraints need to be considered in the process of developing reasonable solutions. Below shows the details of the three constraints in networking systems:

\begin{itemize}
       \item Spatial constraint: A network is composed of a series of entities.
        \item Temporal constraint: Network tasks are sequential operations performed on network entities. 
        \item Causal constraint: Performing an action causes state changes of the network.
    \end{itemize} 

To incorporate these constraints in the design of language systems for zero touch networks, there needs to be a way to construct these constraints as grammar rules of the network language, so that the language derived from the grammar rules can be naturally adaptable and executable in the zero touch network. These three constraints ensure that a language conforms to the semantics of the network domain. 

In this article, we explore the zero touch network from the perspective of language-like systems that 1) reveal the three constraints in the form of a language system, and use Spatial-Temporal-Causal AND-OR Graph (STC-AOG) as the grammatical structure of the language system; 2) given a network task, the Spatial-Temporal-Causal Parsing Graph (STC-PG) can be obtained by parsing the STC-AOG, so as to know what the network task is, how to perform it, and what network intents will be affected; 3) provide network operators with an intuitive understanding of the current network using the translation of STC-AOG or STC-PG into first-order languages.

\section{State-of-the-Art}
The overall goal of zero touch networks is to enable networks to become more autonomous and allow network operators to confidently entrust them with complex, tedious tasks. This fascinating vision has attracted continued attention from the research community, standards organizations and network providers. European Telecommunications Standards Institute (ETSI) established the Zero touch network and Service Management Industry Specification Group (ZSM ISG). Its goal is to eventually automate 100\% of network operations processes. The group has published 10 standards to define an end-to-end operational framework for the future of intelligent networks, enabling networks to be agile, efficient and automated~\cite{953ZSM9883}. Huawei and Juniper  have each released their zero touch network related products, Autonomous Driving Networks (ADN)~\cite{adn} and Self-Driving Networks~\cite{self-driving}, respectively. The ADN has produced a three-layered open architecture delivering intelligence for networks, which are cloud intelligence for data training as well as model generation and optimization, network intelligence for service intent automation, and network element intelligence for providing short period awareness analysis and inference capabilities. The self-driving network aims to ease the heavy lifting of network administrators by self-configure, monitor, manage, correct, defend and analyze with little human intervention.

 Artificial Intelligence (AI) powered by machine learning  and deep learning technologies has been considered as a key enabler for zero touch networks. Alongside standards organisations and network providers, research community is  progressing AI initiatives looking at autonomous networks. Chergui et al. \cite{9749186} proposed a framework with a distributed and AI-driven management and orchestration system for large-scale deployment of network slicing in the sixth-generation network (6G). From the perspective of automata learning, Luque-Schempp et al. \cite{9785739} built an intelligent controller based on learning from the live network, which solves the problem of dynamically predicting and applying the appropriate 5G non-public networks configuration to meet the requirements of the time sensitive network traffic. The authors in \cite{9687472} introduced a zero touch 6G large-scale network slice analysis engine based on statistical federated learning, which performs slice-level resource prediction through offline learning. Shaghaghi et al. \cite{9539883} proposed a zero touch, deep reinforcement learning based proactive failure recovery framework for stateful network functions virtualization enabled networks.

The above work has proposed solutions in various scenarios, but lacks a unified intelligent expression model to express the intent of network tasks, what the task is to do and how to do it. This article manages to make up for this lack from the perspective of language-like systems.

\section{The Hierarchical Language Framework}
In this section, we discuss three important issues of the hierarchical language framework, as shown by the three light green boxes in Fig. \ref{fig:Program overview}. The first (bottom) one is the language system. This part mainly discusses the grammar, symbols and semantics, which are the building blocks of the language system. The second (middle) one is to discuss how to use the language system to construct a unified framework for understanding networking systems. The last (top) one discusses the logical description produced by the network language system to provide network engineers/operators with an understandable description of the current network. It is an interface between the network language system and the human language system.

\subsection{The network language system}
\subsubsection{Grammar}

Composability and reconfigurability are natural properties of network systems. Composability refers to network concepts such as network tasks and network environments, which are composed of multiple sub-steps or sub-components. Reconfigurability means that each sub-step or sub-component has multiple options to choose from. These two natural properties can be covered by AND and OR logic operations respectively. Analogous to linguistics, AND and OR logic operations can be thought of as a representation that encodes a ``grammar''. We take the AND-OR graph (AOG), a directed tree form, as a grammar structure to build the network language system. 
The AOG is a stochastic context-free grammar which is defined as a 5-tuple $<\Omega, N, S, R, P>$.  $\Omega$ is a set of terminal nodes that are not decomposable. $N$ is a set of non-terminal nodes representing decomposable patterns, which can be divided into two disjoint sets: AND-nodes $N^{AND}$ and OR-nodes $N^{OR}$. The AND-node stands for composability, which acts as a parent node and connects with its decomposed child elements. The OR-node stands for reconfigurability. It acts as a parent node and connects to its selectable child elements.
$S$ is a start symbol that belongs to $N$. $R$ is a set of grammar
rules, each of which represents the generation from a non-terminal node to a set of non-terminal or terminal nodes.  The set of grammar rules $R$ is divided into two disjoint sets: AND-rules and OR-rules. $P$ is the set of probabilities assigned to the OR-rules.

Grammar captures the constrains in the networking system. Networking systems have common constrains of organization in spatial, temporal and causal dimensions. For example, a network task is a series of sub-steps distributed over time. The network environment is composed of multiple network functional entities in a spatial layout. In addition, the sub-steps of a network task can cause changes in the state of the entities due to causality. Therefore, the grammar of the network can be designed to represent in terms of spatial, temporal, and causal, so as to obtain the constrains of networking systems more accurately. The spatial-temporal-causal AOG (STC-AOG) which consists of spatial AOG (S-AOG), temporal AOG (T-AOG) and causal AOG (C-AOG), is devised to represent the grammar.

\subsubsection{Symbolization}

 In order to allow AOG to integrate with the network, the terminal node is the symbolization of a network element. In S-AOG, termination nodes are symbolic representations of network entities that exist in physical or virtual networks, including packets, switches, links, storage, just to name a few. In T-AOG, terminal nodes are symbolic representations of micro-actions of network tasks that are applied to network entities, such as adding, deleting, modifying  or querying the storage. In C-AOG, a termination node is an attribute of a network entity that can be changed by network tasks.

By symbolizing network entities, micro-actions, and attributes, analogous to human language expressions, these symbols are arranged to form ``sentences" to express network tasks. For example, a simple task of creating a forwarding tunnel can be expressed as three operations: encapsulation, routing, decapsulation. Encapsulation is to add tunnel information into the packets header. Routing is to forward packets according to the encapsulated information, and decapsulation is to delete the tunnel information of the packet when it reaches the destination. We enumerate several  symbols of common micro-actions as follows: 

\begin{itemize}
       \item $A_{1}$: Delete a specific field in the packet.
        \item $A_{2}$: Add fields at specific locations in the packet.
        \item $A_{3}$: Check a specific field in the packet.
        \item $A_{4}$: Select a port to forward the packet. 
    \end{itemize}

Each operation can be further decomposed into symbols representing micro-actions:

\paragraph{Encapsulation}
$A_{3}A_{2}$. $A_{3}$ is to focus on a position of the packet, and $A_{2}$ is to add fields at specific locations of the packet.

\paragraph{Routing}
$A_{3}A_{4}$. $A_{3}$ is to check the destination address field position, and $A_{4}$ is to move the packet to another location.

\paragraph{Decapsulation}
$A_{3}A_{1}$. $A_{3}$ is to focus on a position of the packet, and $A_{1}$ is to delete fields at the specific locations of a packet.

Therefore, the task can be expressed as a sequence of symbols from the temporal perspective: $A_{3}A_{2} A_{3}A_{4} A_{3}A_{1}$.

\subsubsection{The learning of the S-AOG and T-AOG}
\label{sec:network_language_system}

With only S-AOG, T-AOG and C-AOG terminal nodes, it is difficult to obtain their joint representations. In order to express S-AOG, T-AOG, and C-AOG jointly as STC-AOG, it is necessary to obtain the non-terminal nodes of AOG. With the ground symbol as the termination node of the three AOGs, the non-terminal node can be obtained by recursive induction starting from the termination node. There are two principles in the process of a recursive induction. One is that the tree structure should be as compact as possible, so that non-terminal nodes can represent more abstract semantics. The other is to maximize the likelihood probability of the data given the tree structure, so that the tree structure can accurately explain the data. From the perspective of Bayesian inference, the former is the prior probability, and the latter is the likelihood probability. The essence of inferring AOG from data is to increase the posterior probability which is a multiplication value of these two probabilities. Next we briefly discuss the process of initializing and iteratively generating the S-AOG and T-AOG.

\begin{itemize}
       \item Step1: Define an initial grammar that exactly generates the training dataset;
        \item Step2: Add new grammar fragments to grammars rooted at a new non-terminal node, and the new grammar rules will specify how the new non-terminal node generates one or more configurations of an existing terminal or non-terminal node;
        \item Step3: Reduce each training sample with new grammar rules;
        \item Step4: Steps 2 and 3 are repeated continuously in a bottom-up iterative manner, introducing new intermediate non-terminal nodes between terminal nodes and top-level non-terminal nodes to generalize the grammar.
    \end{itemize}

In a nutshell, the method starts with a simple initial grammar and iteratively inserts new fragments into the grammar to optimize its posterior probability. The algorithm terminates until no more grammar fragments are found to increase the posterior probability of the grammar.

\subsubsection{The learning of the C-AOG}
Unlike S-AOG and T-AOG, which are constructed in a bottom-up manner from terminal nodes to the root node as described in the previous subsection \ref{sec:network_language_system}, C-AOG is learned in a top-down manner. As shown in the top box in Fig. \ref{fig:unified}, the non-terminal nodes of the C-AOG are all measurable features of a network.  The root node of C-AOG is a feature item that represents the intent, such as the transmission delay of data uploading to a cloud center, the server throughput of live streaming and other measurement features. The reason of having the feature of the root represent intents is because they are often the metrics that network 
operators use to measure the performance of network systems. The AND logic between non-terminal nodes is determined by the semantic relationship between features. Semantics can be obtained by self-supervised machine learning methods, as shown in Fig. \ref{fig:Semantic-Self-Supervised}. The method firstly symbolizes all the measured features (shown in the bottom circles), including the features representing the intent (shown in the green circle), and converts each symbol into a compact embedding vector through the operation of the embedding layer. The transformer's encoder model converts the embedding vector into a semantic vector in the semantic space, as shown by the rectangular bars in Fig. \ref{fig:Semantic-Self-Supervised}. In order to obtain the AND logical relationship between the intent feature and the other features, the semantic vectors of both types of features calculate the cosine similarity. Each cosine similarity is multiplied by the value of the corresponding feature to filter irrelevant information. Finally, a discriminative model is used to calculate the consistency between the filtered information and the intent value. Through this method, the model can obtain the features with the high semantic similarity and the intent features under the condition of high consistency. The intent feature serves as the root AND node, and the semantically similar features serve as its child nodes. The C-AOG is thus initially constructed. This method can be used iteratively to build a lower-level structure from top to bottom, and finally connects the possible values or value ranges of each feature with an OR logical relationship.

\subsubsection{The parameter learning of the OR edge}
The OR edge of an AOG is defined as an edge connecting an OR node to its children. The edge has a probability as the parameter, which is defined as the frequency of occurrence of the child nodes connected by the edge, that is, the ratio of the number of occurrences of this child node to the number of occurrences of all sibling nodes.

\subsection{A unified framework for understanding networking systems}
\label{sec:unified_framework}

\begin{figure*}[htb]
	\centering
	\includegraphics[width=1\linewidth]{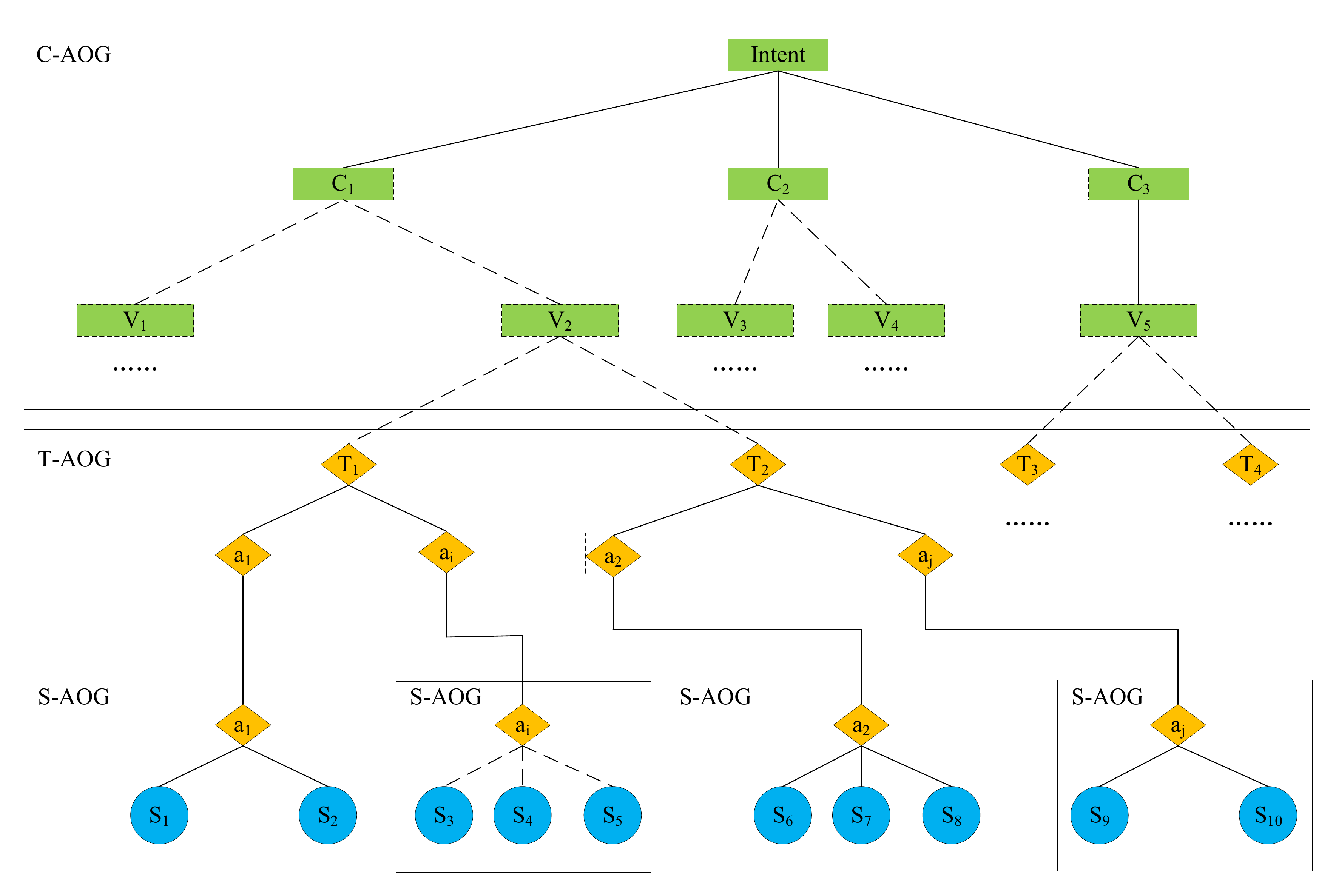}
	\caption{A unified framework for understanding networking systems. }
	\label{fig:unified}
\end{figure*}

This section discusses the fusion of S-AOG, T-AOG, and C-AOG into STC-AOG, in order to express the entire networking system within a unified framework driven by intents. As shown  in Fig. \ref{fig:unified}, from bottom to top, the root node of S-AOG is connected to the terminal node of T-AOG, since each micro-action of network operations requires interaction with network entities. The parent node of the terminal node of C-AOG represents the state symbol. If the state is discrete, the terminal node (child node) represents the specific discrete value. If the state is continuous, the terminal node represents the range of values. The root node of the T-AOG is connected to the terminal node of the C-AOG, because the network operations are the cause of network state changes.  In this way, the three independent S/T/C-AOGs are fused into an integral STC-AOG. 

The link between S-AOG and T-AOG can be obtained from network data through machine learning. The machine learning methods learn the conditional probabilities of micro-actions given the representation of network entities, and connect network entities to the most likely micro-actions.  The link between C-AOG and T-AOG is to connect the C-AOG terminal node with the network operation with the highest probability that is calculated by the conditional distribution of each state value or the value interval under a given network operation.

\begin{figure}[htb]
	\centering
	\includegraphics[width=1\linewidth]{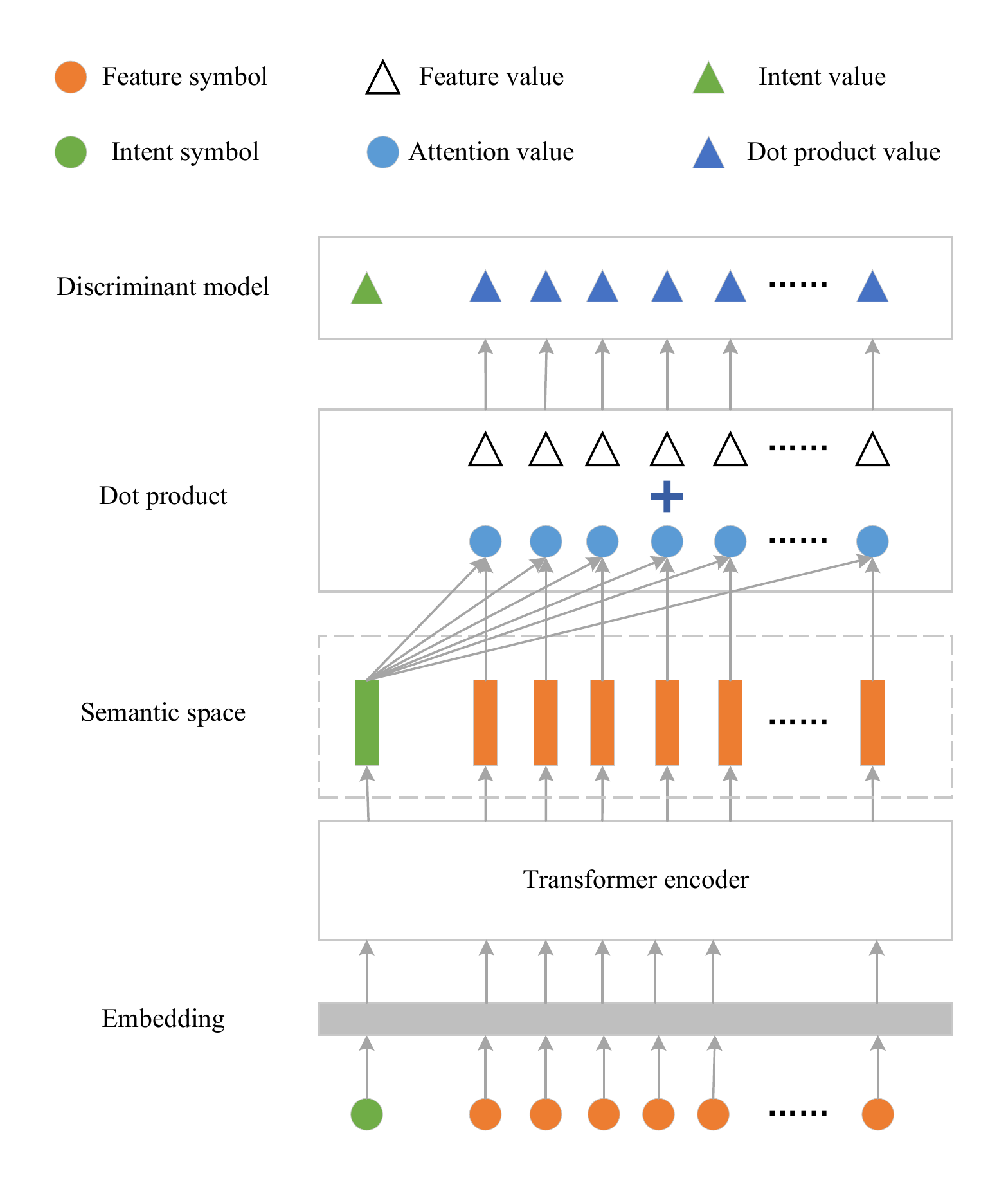}
	\caption{A semantic self-supervised learning method for non-terminal nodes of C-AOG. }
	\label{fig:Semantic-Self-Supervised}
\end{figure}

The STC-AOG is a consistent representation framework for large-scale network knowledge at all levels of abstraction. STC-AOG contains all valid parsing graphs (PG), which are the interpretations of a particular network by deterministically selecting branches of OR nodes in the STC-AOG. The understanding of networking systems is considered as the problem of parsing the STC-AOG to obtain the semantic meaning of a network. The process of interpretation  refers to starting from the topmost C-AOG intent root node and finding a path through T-AOG and S-AOG in turn. The path can be solved using the Viterbi algorithm~\cite{9581213858}, which represents the path with the highest likelihood probability. It interprets the network semantics of the collected data.

\subsection{The logical description}
The logical description is expressed as a logical sentence based on network semantics. It provides a human-readable interface for intuitively understanding network systems. The main information of STC-AOG contains network entities, the behavior of the entity, the state changes by the behavior, and the intention affected by the state changes.  In order to facilitate the complete expression of these information, the form of logic description adopts the first order logic  (FOL) due to its unique advantages.

The form of FOL consists of four components: individual domain, predicate, logical symbol, and quantifier. Predicates and individual domain are used to represent parent and child nodes, respectively. The logical symbols $\wedge$ and $\vee$ represent the AND and OR relationship of AOG, respectively. In addition, the first-order logic has the ability to express quantifiers which refer to the number of individuals. There are universal quantifiers, marked with $\forall$, and existential quantifiers, marked with $\exists$. The $\forall$ quantifier is used to cover all individuals satisfying the predicate. The $\exists$ quantifiers are used to describe some individuals that satisfy the predicates. Quantifiers are used to describe the connection between C-AOG and T-AOG. Existential quantifiers denote the OR concatenation, representing a change in states e.g. due to a particular operation. Universal quantifiers represent the AND connect, representing that state changes are always due to certain specific operations.

We take Fig. \ref{fig:unified} as an example to illustrate how to describe the STC-AOG with the first-order logic. For the sake of simplicity, the ellipsis part in the figure is not described. According to the STC-AOG from top to bottom, there are five logical sentences used to describe the information. The first sentence is about the information of C-AOG in the fragment. The second and third sentences are about the relationship between T-AOG and C-COG, and the last two sentences are about the relationship between S-AOG and T-AOG.

\begin{enumerate}[]
\item $ ( C_1(V_1) \vee C_1(V_2) ) \wedge ( C_2(V_3) \vee C_2(V_4) ) \wedge C_3(V_5) \to Intent $
\item $ {\exists} T_1  {\exists} T_2\ (V_2(T_1) \vee V_2(T_2) )$
\item $ {\exists} T_3  {\exists} T_4\ (V_5(T_3) \vee V_5(T_4) )$

\item $a_1(S_1, S_2) \wedge (a_i(S_3)  \vee a_i(S_4)  \vee a_i(S_5)) \to T_1$
\item $a_2(S_6, S_7, S_8) \wedge (a_j(S_9, S_{10})) \to T_2$
\end{enumerate}

The interpretation process is to deterministically select a child node for the OR node in the STC-AOG, thereby obtaining the STC-PG. For example, if the $V_2$, $V_3$, $T_1$, $T_3$ and $S_4$ are selected, the STC-PG can be described by the first-order logic as follows:

\begin{enumerate}[]
\item $   C_1(V_2)  \wedge  C_2(V_3)  \wedge C_3(V_5) \to Intent $
\item $ {\forall} T_1 {\forall} T_3 ~ (V_2(T_1) \wedge V_5(T_3)) $
\item $a_1(S_1, S_2) \wedge  a_i(S_4)  \to T_1$

\end{enumerate}

\section{A Case Study} \label{usecase}

\iffalse
\begin{figure*}[ht]
	\centering
	\includegraphics[width=1.0
	\linewidth]{"serviceTime_intent.png"}
	\caption{Comparison of service time between human semantic model and network semantic model}
	\label{fig:serviceTime_intent}
\end{figure*}
\fi

\begin{figure*}[p]
	\centering
	\includegraphics[width=1\linewidth]{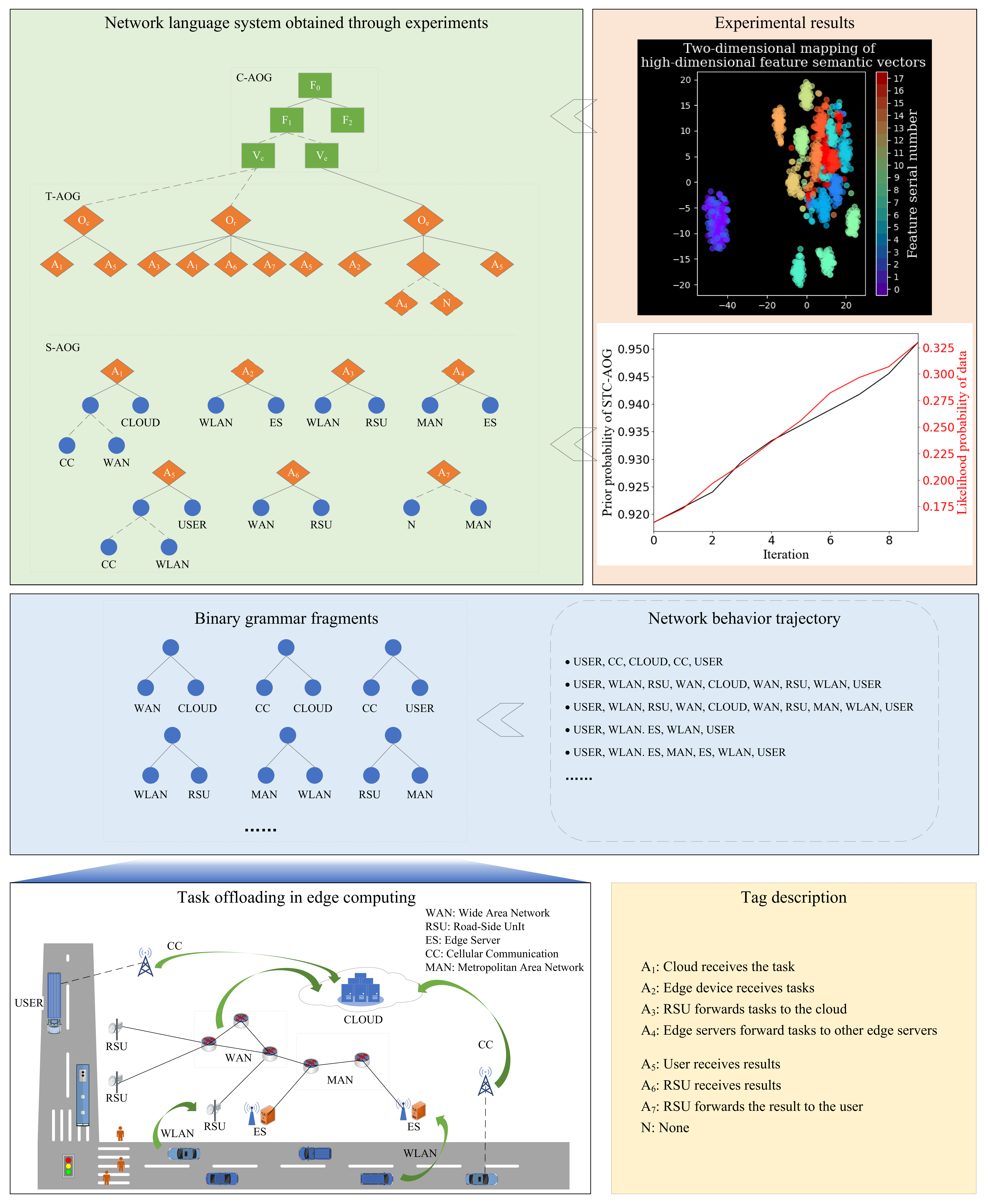}
	\caption{A case study of the proposed language framework }
	\label{fig:Case_study}
\end{figure*}

\begin{figure*}[ht]
	\centering
	\includegraphics[width=1.0
	\linewidth]{"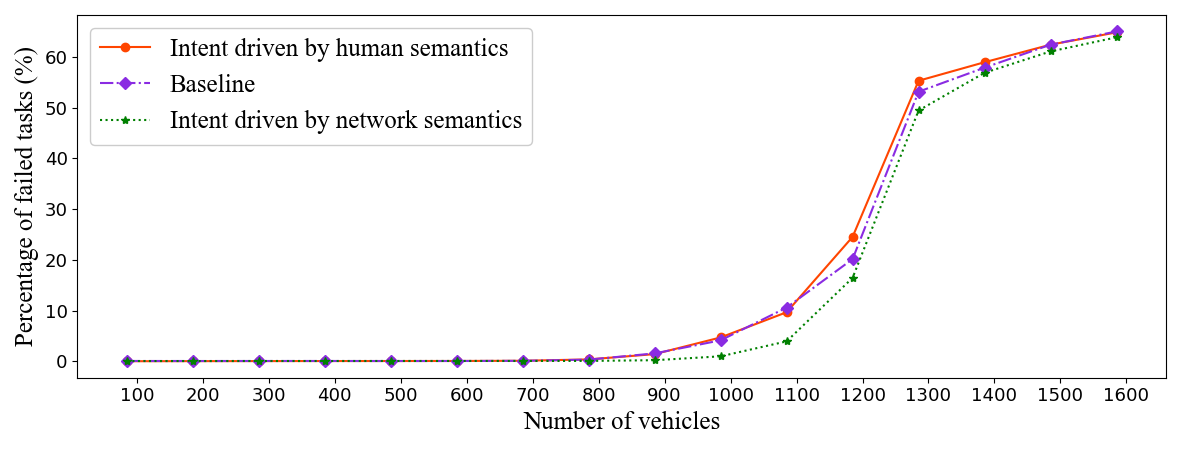"}
	\caption{Comparison of task failure rates between human semantic models and network semantic model}
	\label{fig:completion_intent}
\end{figure*}

To verify the capability and feasibility of the proposed the hierarchical language framework, we present a networking scenario as an example: task offloading in edge computing. Task offloading is a common application in the field of Internet of Things and edge computing. When the user equipment cannot provide sufficient computing resources, the task is offloaded to a suitable edge server or a cloud center for auxiliary service. 

The considered scenario is shown in the bottom part of Fig. \ref{fig:Case_study}. The road-side unit (RSU) and edge server (ES) are deployed at the road side. ES provides nearby computing services for vehicular applications, and RSU assists vehicular applications in terms of forwarding tasks to the cloud through the WAN. Considering the movement of vehicles, all RSUs and ESs are connected through the MAN to achieve roaming. In addition, vehicular applications can also communicate with roadside base stations and offload tasks to the cloud via the cellular mobile network. Therefore, a task has three options for offloading: the first one is to offload to ES; the second one is to offload to the cloud through RSU; and the last one is to offload to the cloud through cellular mobile networks. We deploy probes on the ports of network entities to collect path data for task offloading. The path data is shown as the network behavior trajectory in the blue box in Fig. \ref{fig:Case_study}.  Each record represents a path. Paths are represented by strings of probe symbols, which denote the flow of traffic through these entities. According to the path data, all possible combinations of the two probe symbols are formed to construct a set of binary grammar fragments, as shown on the left side of the blue box in Fig. \ref{fig:Case_study}. These fragments are the basic blocks to build  grammatical structures of the S-AOG and T-AOG. The root node of the fragment represents a possible micro-action, and the child nodes 
denote the entities involved in the micro-action. Each record is treated as a complete operation that contains multiple micro-actions.

The learning of STC-AOG is to obtain a compact grammar structure that can explain the path data. In the experiment, the normalized exponential function is used as the prior probability of the grammar structure, which is proportional to the negative of the number of nodes of the grammar structure. Hence, it serves as an indicator of compactness. The more compact the grammar structure, the larger the prior probability. In order to avoid blind pursuit of compactness, which reduces the ability of grammar to explain data, likelihood probability is used to describe the probability of data under a given grammar structure. As shown in the lower figure of the experimental results box (the pink box) in Fig. \ref{fig:Case_study}, as the number of inductive iterations increases, the prior and likelihood 
rise first and then stabilize, indicating that the algorithm can find an optimal compact structure  that conforms to the path data. The grammar structure is shown in the T-AOG and S-AOG boxes in the figure. From the experimental results, it can be seen that the number of fragments of S-AOG is reduced to 7 (compared to 28 fragments of the binary grammar fragments). The $A_1$, $A_5$, $A_7$ segments in S-AOG have OR nodes, which increase the compactness of S-AOG. The learned T-AOG has three operations, $O_c$, $O_r$ and $O_e$, which respectively indicate that computing tasks are offloaded to the cloud through the cellular network, offloaded to the cloud through the RSU, and directly offloaded to the ES. These three operations cause the $F_1$ feature to change, where $V_c$ means the task is executed in the cloud, and $V_e$ represents the task is executed at the edge.

\begin{table*}[!htbp]
    \caption{The 18 features collected in the experiment and their associated descriptions}
	\centering
	\begin{tabular}{|l|l|}
		\hline
		 Symbols& Remark\\ \hline
	     $F_{0}$     & Indicates the task completion status.\\ \hline
		 $F_{1}$    &Indicates the destination of the task to be offloaded.\\ \hline
		 $F_{2}$  &Indicates the offloading service time.\\ \hline
		 $F_{3}$ &Indicates the location of the vehicle. \\ \hline
		 $F_{4}$     & Indicates the file size of the task.\\ \hline
		 $F_{5}$    &Indicates the size of the result of the task.\\ \hline
		 $F_{6}$  &Indicates the number of instructions for the task.\\ \hline
		 $F_{7}$ &Indicates the upload delay of the WAN network. \\ \hline
		 $F_{8}$     & Indicates the download delay of the WAN network.\\ \hline
		 $F_{9}$    &Indicates the upload delay of the cellular network.\\ \hline
		 $F_{10}$  &Indicates the download delay of the cellular network.\\ \hline
		 $F_{11}$ &Indicates the upload delay of the WLAN network. \\ \hline
		 $F_{12}$     & Indicates the download delay of the WLAN network.\\ \hline
		 $F_{13}$    &Indicates the average utilization rate of the edge server.\\ \hline
		 $F_{14}$  &Indicates the average utilization rate of the cloud server.\\ \hline
		 $F_{15}$ &Indicates the number of tasks offloaded to the edge server recently. \\ \hline
		 $F_{16}$     & Indicates the number of tasks offloaded to the cloud via RSU in the recent period.\\ \hline
		 $F_{17}$    &Indicates the number of tasks offloaded to the cloud via the cellular network in the recent period.\\ \hline
		 
	\end{tabular}
	\label{tab:features}
\end{table*}

In the experiment, we collect 18 features and the corresponding meaning of these features are shown in Table \ref{tab:features}, of which the feature $F_0$ is the intent feature.  It indicates that network operators want to maximize the success rate of task offloading (task execution completes and successfully returns to the user device). The upper figure of the experimental results box (the pink box) in Fig. \ref{fig:Case_study} is the visualization of the 18 features in the semantic space. The color bar on the right is the color represented by feature $F_0$ to feature $F_{17}$, respectively. It can be seen that the color of the feature $F_0$  is fused with the colors of the feature $F_1$ and the feature $F_2$. This reveals that the semantics of these three features are highly correlated so that  $F_0$  is the root node, and $F_1$  and $F_2$  are its child nodes in C-AOG. Based on the features semantic learning for C-AOG and the  prior and likelihood probabilities learning for S-AOG and T-AOG, the experiment yields a grammar structure STC-AOG of computational offloading tasks for edge computing scenarios. 

For each path data, STC-PG can be parsed from the learned STC-AOG, and then translated into first-order logic. For example, the record \{USER, CC, CLOUD, CC, USER\} can be interpreted into the first-order logic as follow:

\begin{enumerate}[]
\item $   F_1(V_c) \wedge F_2(*)  \to Intent $
\item $ {\forall} O_c  ~ (V_c(O_c) ) $
\item $A_1(CC, CLOUD) \wedge  A_5(CC, USER)  \to O_c$

\end{enumerate}

This first-order logic statement visualizes the relationship between the intent (what to do) and a specific task (how to do).  The first sentence shows that the value of $F_1$ changing to $V_c$ will affect the realization of the intent. The second sentence further indicates that only this operation is the reason for changing the state of feature $F_1$ to $V_c$. The last sentence explains in detail how the operation is performed. The micro-action $A_1$ is executed on $CC$ and $CLOUD$, and the micro-action $A_5$ is executed on $CC$ and $USER$. These two micro-actions are executed in sequence to complete $O_c$ operations. It means the task is offloaded to the cloud and backtracked to the user device. 

To verify the capability of the proposed language system in network optimization, we conduct additional experiments. According to the previous experiments, the intent feature $F_0$ is related to the  feature $F_1$. Three operations $O_c$, $O_r$ and $O_e$ can affect the value of the $F_1$. We utilize the previous semantic learning model as shown in Fig. \ref{fig:Semantic-Self-Supervised} to learn the semantic relationship between the intent feature $F_0$  and the entities involved in $O_c$, $O_r$ and $O_e$ operations (an entity is represented by its attributes). We obtain the entity attributes that are most relevant to the intent feature. Based on this, we design a decision model, where the input is an entity attribute that is semantically strongly related to the intent feature, and the output is the probability distribution of the three operations. The operation with the highest probability indicates that the model predicts that it is the most likely operation to achieve the optimization intent.

The experimental results are shown in Fig. \ref{fig:completion_intent}, and the purple line is the performance of the stochastic decision model on the intent of the task failure rate. The decision models of the red and green lines use the data generated by the stochastic decision model as training data. The input of the decision model of the red line is the features designed based on human prior knowledge, and that of the decision model of the green line is the feature that is strongly related to the intent learned by our proposed language system. Experimental results show that our decision model performs optimally. It can correct the bias introduced by the artificially designed model, so that the input features of the model are more in line with the network intent.

\section{Research Challenges and Open Issues} \label{Challenges and Open issues}
Many studies have been reported for zero touch networks. In this article, our solution is proposed by building a language system which is specifically designed for the networking domain. In this section, we discuss research
challenges and open issues which can guide the further research on this topic.

\subsection{Sustainable evolution of network intelligence} \label{challenge1}
Zero touch network requires minimal human involvement in the operating of the network. It is therefore required that the form of network intelligence can continue to evolve with the changing network states. From the initial stage, it may only express a small amount of knowledge, and then continue to evolve to express more abundant network knowledge. Embedding a sustainable evolving network intelligence into the design of zero touch networks is a challenge.  The characteristics of network applications vary over time, and the network architecture is complex and diverse. To allow knowledge representation to cover most applications and network components, knowledge representation needs to build in the commonalities behind these complexities.  In addition, with the commonality as the basis for constructing knowledge representation, the knowledge representation model can sensitively perceive changes in network components or network tasks, and accurately adjust the knowledge structure to conform to the new network state.

\subsection{Alignment of network intelligence with human semantics}
\label{challenge2}
The zero touch network not only requires it to operate in a self-driving manner, but also needs to expose the current state of the network to network operators, so that the situation of the network can be observed. This is even more important in a production environment, where administrators can grasp the status of the network in real time to ensure normal production. This challenge boils down to the problem of aligning network intelligence with human semantics. This is because the form of intelligence embedded in the network may be a vector or symbolic representation. The semantics of these vectors or symbols need to be aligned with human semantics in order to be understood by human administrators. Especially when these vectors or symbols are learned directly from the data, their semantics are directly induced from the data. There is a gap between data-induced semantics and human semantics. Guaranteeing free of alignment issues  to close that gap is also worth of investigation.

\subsection{Endowing zero touch networks with value theory}
\label{challenge3}
An important mission of zero touch networks is to keep the network system running in an optimal state. This requires zero touch networks to have the ability to judge value so as to understand what is optimal. It is not trivial to formalize the theory of value and embed it into an intelligent representation model so that the model can make decisions to take optimal actions consistent with the value for the network. This is because the value pursuit of the network is diverse. The value pursuit of each task only focuses on itself, but the overall pursuit of the network is fairness. The decision made by the model may conform to the value of a certain network application, such as reducing the transmission delay of the application, but it may not be able to meet the value pursuit of the entire network at this time, such as requiring the performance of all applications to be balanced. In addition, the value theory embedded in the intelligent representation model is the basis for ensuring that the behavior of zero touch networks is credible.

\subsection{Using small-scale data to learn and optimize}
\label{challenge4}
Learning and optimization of networks usually require a large amount of data.  However, insufficient data collection or missing data are often encountered in network management.  One way to achieve optimization based on small-scale of data is to collect the specific important feature data based on prior knowledge and the intent of the optimization. This is an urgent problem to be solved in the separation of data collection and model optimization. It is a real obstacle to the realization of closed loop controls of zero touch networks. Building an intelligent representation model embedded in the closed loop control creates opportunities for optimization based on small data scales, since the representation contains a lot of prior knowledge. Another way is to study the ability of intelligent representation models to draw inferences from one another, namely one-shot or few-shot learning. It can quickly generalize a previously trained model for new tasks with only a few samples.

\section{Conclusion}
This article proposed a hierarchical language  framework for zero touch networks. In the language system of the framework, the spatial, temporal and causal constraints endogenously in the network system are expressed explicitly by the grammar structure of STC-AOG. This article discussed the initializing and learning the  STC-AOG, the parsing of STC-AOG to obtain STC-PG, and the transformation of hierarchical STC-AOG into a flattened symbol string in the form of first-order logic. Given a task, the zero touch network knows what the network task is, how to perform it, and what intents will be affected with the help of the network language system. In addition, 
symbols strings induced from the network language system provides network operators with an intuitive understanding of the network. A case study
was developed to demonstrate the feasibility of the proposed
hierarchical language  framework for zero touch networks by performing
the scenario of task offloading in edge computing. At last, we provided a list of research challenges and open issues that can be useful to the community
for carrying out further research.

\iffalse
\section*{Acknowledgment}

The authors would like to thank...
\fi

\ifCLASSOPTIONcaptionsoff
  \newpage
\fi

\bibliographystyle{IEEEtran}
\bibliography{ref}

\begin{IEEEbiographynophoto}{Guozhi Lin}
is now a Ph.D. candidate of the University of Chinese Academy of Sciences. He received the B.Sc. degree in computer science and technology from South China Normal University in 2016. His research interests are mainly in intent-based networks, network automation and reinforcement learning.
\end{IEEEbiographynophoto}

\begin{IEEEbiographynophoto}{Jingguo Ge}
is currently a Professor with the Institute of Information Engineering, Chinese Academy of Sciences, and a Professor with the School of Cyber Security, University of Chinese Academy of Sciences. He received the Ph.D. degree in computer system architecture. His research focuses on computer network architecture and cyber security.
\end{IEEEbiographynophoto}

\begin{IEEEbiographynophoto}{Yulei Wu}
[Senior Member] is a Senior Lecturer with the Department of Computer Science, College of Engineering, Mathematics and Physical Sciences, University of Exeter, UK. He received the Ph.D. degree in Computing and Mathematics from the University of Bradford in 2010. His research interests include networking, IoT, edge intelligence, privacy and trust, and AI and ethics.
\end{IEEEbiographynophoto}

\end{document}